\documentclass[conference]{IEEEtran}
\IEEEoverridecommandlockouts
\usepackage{cite}
\usepackage{amsmath,amssymb,amsfonts}
\usepackage{algorithmic}
\usepackage{graphicx}
\graphicspath{{images/}}
\DeclareGraphicsExtensions{.png}
\usepackage{textcomp}
\usepackage{xcolor}
\usepackage{subfig}
\usepackage{multirow}

\usepackage{tikz}
\usepackage{textcomp}
\usepackage{hyperref}
\usepackage{lipsum}

\def\BibTeX{{\rm B\kern-.05em{\sc i\kern-.025em b}\kern-.08em
    T\kern-.1667em\lower.7ex\hbox{E}\kern-.125emX}}

\newcommand\copyrighttext{%
  \footnotesize \textcopyright 2019 IEEE. Personal use of this material is permitted. Permission from IEEE must be obtained for all other uses, in any current or future media, including reprinting/republishing this material for advertising or promotional purposes, creating new collective works, for resale or redistribution to servers or lists, or reuse of any copyrighted component of this work in other works.
Accepted paper for 2019 8th International Conference on Affective Computing and Intelligent Interaction Workshops and Demos (ACIIW). For citation and published version DOI, see below:\\
J. O’Dwyer, “Speech, Head, and Eye-based Cues for Continuous Affect Prediction,” in 2019 8th International Conference on Affective Computing and Intelligent Interaction Workshops and Demos (ACIIW), IEEE, Sep. 2019, pp. 16–20, isbn: 978-1-7281-3891-6. doi: \href{https://doi.org/10.1109/ACIIW.2019.8925042}{10.1109/ACIIW.2019.8925042}
}
\newcommand\copyrightnotice{%
\begin{tikzpicture}[remember picture,overlay]
\node[anchor=south,yshift=10pt] at (current page.south) {\fbox{\parbox{\dimexpr\textwidth-\fboxsep-\fboxrule\relax}{\copyrighttext}}};
\end{tikzpicture}%
}

\begin{document}

\title{Speech, Head, and Eye-based Cues for Continuous Affect Prediction
\thanks{This work was supported by the Irish Research Council (Grant Nos. GOIPG/2016/1572, GOIPG/2018/2030.}
}
\author{\IEEEauthorblockN{Jonny O'Dwyer}
\IEEEauthorblockA{\textit{Department of Computer \& Software Engineering} \\
\textit{Athlone Institute of Technology}\\
Athlone, Ireland \\
j.odwyer@research.ait.ie}
}

\maketitle
\copyrightnotice

\begin{abstract}
Continuous affect prediction involves the discrete time-continuous regression of affect dimensions. Dimensions to be predicted often include arousal and valence. Continuous affect prediction researchers are now embracing multimodal model input. This provides motivation for researchers to investigate previously unexplored affective cues. Speech-based cues have traditionally received the most attention for affect prediction, however, non-verbal inputs have significant potential to increase the performance of affective computing systems and in addition, allow affect modelling in the absence of speech. However, non-verbal inputs that have received little attention for continuous affect prediction include eye and head-based cues. The eyes are involved in emotion displays and perception while head-based cues have been shown to contribute to emotion conveyance and perception. Additionally, these cues can be estimated non-invasively from video, using modern computer vision tools. This work exploits this gap by comprehensively investigating head and eye-based features and their combination with speech for continuous affect prediction. Hand-crafted, automatically generated and CNN-learned features from these modalities will be investigated for continuous affect prediction. The highest performing feature sets and feature set combinations will answer how effective these features are for the prediction of an individual’s affective state.
\end{abstract}


\section{Introduction}
Affective computing can be thought of as the use of computers related to human or human-like feelings. Such use of computers can include human emotion or psychopathology recognition based on audio-video data, or synthesising human-like emotional speech for robotics applications. Speech-based affective computing is now well developed. There is nearly thirty years of research related to speech-based affective computing \cite{b1} and 66\% of the world's native language speaking populations are represented by affective speech data sets \cite{b2}. Despite a large body of evidence linking eye and head-based cues to emotion and motivational state conveyance \cite{b3,b4,b5,b6,b7,b8,b9,b10,b11,b12}, the use of these cues is underdeveloped for affective computing purposes.

Based on the identified research opportunity, answering the following research question is the aim of this work: How significant an improvement in the prediction of an individual's affective state can be achieved by processing the combined cues gathered from an individual's speech, head and eyes? Therefore, in this paper, work towards finding appropriate features and showing the usefulness of head and eye-based cues estimated from video and combined with speech for continuous affect prediction is described. The layout of the remainder of this paper is as follows. Related work that informed and inspired this project is provided in Section II. This is followed by presentation of the methodology employed in Section III. Initial results are given along with brief discussion of these results in Section IV. Section V concludes this paper in the form of some final remarks and a future work plan.

\section{Related Work}
Within affective computing, continuous affect prediction involves the discrete time-continuous regression of the affective state of individuals. Features extracted from different modalities such as speech or facial expression are used as input to this process. Advantages of continuous affect prediction include the time-dependent nature of prediction and the complex \textit{N}-dimensional affect representation provided. These advantages can allow temporal gradients of affect to be predicted and enable affect or emotion capture that may be outside that of human verbal description \cite{b13}. Common affect dimensions include arousal and valence, which may be plotted in a 2D circumplex model such as that proposed by Scholsberg \cite{b14} and later refined and demonstrated to resemble a cognitive structure of affect by Russell \cite{b15}. 

Speech-based affective computing is now a well developed field. There are corpora \cite{b13, b16, b17, b18}, feature sets \cite{b18, b20, b21, b22} tools \cite{b23} and repositories \cite{b24} to enable speech-based input for affective computing systems. Including speech in affective computing systems, is a good idea whenever possible, as the performance benefits of using this modality are well established, particularly for continuous arousal prediction for example. Additionally, if a baseline of human-level performance is to be considered, including speech in multimodal systems provides for fairer comparisons as humans clearly have access to both verbal and non-verbal output when forming their affect annotations. An interesting new direction in speech affective computing is end-to-end learning. End-to-end learning was performed in \cite{b25} with concordance correlation coefficient (CCC) scores of 0.686 for arousal and 0.261 for valence on the RECOLA \cite{b17} test set. The input features for \cite{b25} were learned directly from raw speech data using a convolutional neural network (CNN) algorithm and the feature vectors were passed to a bidirectional long short-term memory recurrent neural network (BLSTM-RNN) for continuous affect prediction.

Busso et al. \cite{b10} showed head pose and speech prosody to be strongly linked by objective measures in their work. They additionally carried out a subjective experiment that showed emotion perception changing in the presence of different head motion patterns. While in \cite{b11}, the authors concluded that vocalists' head movements encode emotional information during speech and song, and that observers could identify emotion based on head movement alone in their subjective experiment. Objective and subjective experiments were carried out in \cite{b12} for the purpose of understanding how head motions contribute to the perception of emotion in an utterance. For the objective experiment, 45 dynamic features based on the discrete Fourier transform of yaw, pitch and roll angular head movements, gathered using a head-mounted device, along with a static measure of head pitch, were used as input to machine learning algorithms. The best results achieved were 94\% for neutral, 79\% for sad, 57\% for happiness and 72\% for angry classification. These studies clearly indicate that head motion is important for emotion signalling and discrimination, despite this, no studies exist which have investigated head-based features for continuous affect prediction. 

Eye-based cues have been linked with various emotional displays for a long time \cite{b26} and numerous works suggest various eye signals to elucidate one's affective state. The shared signal hypothesis \cite{b5, b6} suggests that eye gaze is congruent with emotion expressions when the gaze direction matches the underlying motivation to approach or avoid stimuli while pupil size variation occurs during monetary reward or penalty \cite{b27}, while viewing emotionally arousing stimuli \cite{b28, b29}, during reward expectation \cite{b30} and during autonomic nervous system stimulation \cite{b31}. In affective computing, eye-based cues have been employed for emotion recognition \cite{b32, b34}. Decision tree neural network was used in \cite{b34} .affect-level recognition \cite{b16} and psychopathology applications \cite{b37, b38}. Soleymani et al. \cite{b16} achieved good performance using eye features comprised of statistics and power spectral density calculations based on eye gaze, eye blink and pupillometry low-level descriptors (LLDs) as input to Support Vector Machine (SVM). The eye-based features performed best for unimodal affect-level recognition when compared to compared to other physiological measures in their experiments. However, specialised equipment is required for gathering the proposed features in \cite{b16}, in the form of an external Tobii eye gaze recorder, which provides a barrier to researchers in accepting and enhancing the proposed features/approaches.

Following the review of related work it is clear that there is evidence supporting the use of head and eye-based cues as inputs to affective computing systems. Furthermore, modern computer vision tools can estimate these cues from video \cite{b36, b38} which can serve to maximise the impact of these cues for affective computing, if their usefullness can be shown. This serves as motivation for this work. It is hypothesised that head and eye-based cues, estimated from video, can improve the performance of affective computing systems, providing performance benefits when combined with speech and providing affect prediction capabilities in the absence of speech. Therefore, the technical scope of this work includes comprehensively investigating these cues combined with speech for continuous affect prediction.

\section{Methodology}
The methodology employed for this work includes gathering LLDs from video followed by BLSTM-RNN model creation and evaluation based on proposed features and feature combinations input. As part of the feature engineering process, some LLDs are captured from raw head pose, eye-based or visual features, while others must be calculated based on differences in the raw data measurement. For example, pupil dilation is calculated as being equal to 1 based on whether the pupil diameter measurement gathered using OpenFace is larger for a frame when compared to an immediately preceding frame measurement whereas a feature such as the raw eye gaze $x$ coordinate is taken directly as a LLD. The LLD extraction processes is then followed by higher-level features extraction, which are gathered under varying temporal feature windows. This is followed by altering the now commonly accepted affect learning parameter, ground-truth backward time-shift, on the validation set in order to provide better performance of extracted features by taking human annotator lag in providing ratings into account. Confirmatory analysis of features' or feature sets' efficacies are provided by way of test set evaluation. Further details on the methodology are given in this section.

\subsection{Corpus and Training, Validation and Test Partitioning}
The RECOLA \cite{b17} corpus is used as the experimental data set for this work. RECOLA is an affective data set comprised of audio-visual and physiological recordings of subjects cooperating on a task and communicating in French. Arousal and valence annotations, ranging from -1.0 to +1.0, are provided with the set in discrete-continuous-time at a rate of 25 values per second. Each recording in the set is 5 minutes in length. Recordings of 23 subjects available in the set were paritioned into training, validation and test sets with the aim of matching the distributions used in \cite{b34}. Specifically, the training set is comprised of subjects [P16, P17, P19, P21, P23, P26, P30, P65], the validation set includes subjects [P25, P28, P34, P37, P41, P48, P56, P58], and the test set includes subjects [P39, P42, P43, P45, P46, P62, P64].

\subsection{Gathering LLDs}
LLDs comprised of raw head pose and eye-based data are gathered using OpenFace 2.0.6 \cite{b38}. The initial set of raw head-based data includes head location $x$, $y$ and $z$ in camera coordinate millimeters, and head rotation yaw, pitch and roll in world coordinate radians with camera origin. The initial set of eye-based data includes pupil diameter, eye gaze $x$, $y$ radians, eye gaze distance, logical eye blink/closure and eye blink intensity. All of the eye-based raw data are based on world coordinates except eye gaze distance, the one camera coordinate LLD. Briefly, world coordinates are independent of the camera whereas camera coordinates depend on camera location.

\begin{figure}[htbp]
\includegraphics[width = 3.5in]{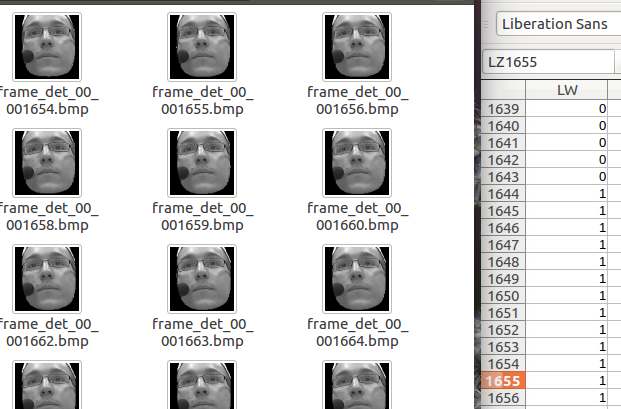}
\caption{Screen capture of human annotation of direct eye gaze.}
\label{figure1}
\end{figure}

A number of additional LLD features are calculated on a frame-wise basis to capture more detailed information on low-level feature dynamics. These features include: displacement (deltas) of the all head features and eye gaze $x$, $y$ radians, and binary true/false features for eye fixation, eye gaze approach, direct gaze, pupil dilation and pupil constriction. All of the additional LLDs are calculated using software resulting from this research while the direct gaze features were annotated by a human observer who judged OpenFace output frames as either direct gaze = 1 (looking at the screen/camera) or 0 (averted gaze away from the screen/camera) as in Fig.~\ref{figure1}.

\subsection{High-level Feature Extraction}
High-level features are extracted using 4, 6 and 8 second temporal windows, moved forward at a rate of 1 frame per interval. Each temporal window is tested using each modality, or combination of modalities, as input to the BLSTM-RNN for training and validation. The high-level features are sometimes preceded by mid-level feature extraction in the form of scale and detail wavelet coefficient features, where 10-order Daubechies wavelet \cite{b39} features are extracted using discrete wavelet transform for as many decomposition levels as possible for a given temporal window. The calculations for the final high-level features include, where appropriate: ratio, time in seconds[min, mean, max, total], min, max, mean, median, quartile 1, quartile 3, skewness, kurtosis, standard deviation, numerous inter-quartile range measurements, linear regression slope, linear regression intercept, RMS, and zero-crossing rate. The high-level features are calculated from static-time (one frame of data) and dynamic-time (frame-wise lagged difference data) measurements of LLDs and the mid-level time-frequency wavelet features for each temporal feature sample window to enhance the data prior to machine learning. Temporal feature windows are denoted $W_{s}$ for the remainder of this work, where ${s}$ indicates the window size in seconds per interval. After gathering feature sets, exploratory data analysis is carried out using feature-to-target relationship calculations.

\subsection{Ground-truth Backward Time-shift}
Following from recent works \cite{b40, b41}, ground-truth annotations provided with the RECOLA \cite{b17} corpus are shifted back-in-time to account for annotator rating time delay. The ground-truth backward time-shift sizes for the experiments range from 0 to 4.4 seconds in steps of 0.2 seconds. These are referred to as $D_{s}$ for the remainder of this work, where ${s}$ indicates the delay in seconds applied to ground-truth annotations prior to concatenation with input features.

\subsection{Feature Selection}
In order to try to achieve the best features sets from the modalities, feature selection is applied in order to remove redundant or weak features inadvertently generated during early feature engineering. The feature selection approach taken in this work follows a simple approach of mutual information (MI) estimation to regression target-based filtering. MI is ``the amount of information that one random variable contains about another random variable'' [40, p.18] and it provides information on the nonlinear relationship between input features and target variables. Features under MI thresholds of 0.1, 0.15 or 0.2 are removed from features sets for experimental evaluation as these features are deemed independent of arousal or valence and therefore poor predictors.

\subsection{BLSTM-RNN Training and Evaluation}
BLSTM-RNN is used in this work to train models for feature set appraisal. The training method largely follows that of Ringeval et al. \cite{b43}. Single-task models are trained using BLSTM-RNN with 2 hidden layers, each with 40 and 30 nodes respectively, with a sum-of-squared-errors (SSE) objective function using the CUDA RecurREnt Neural Network Toolkit \cite{b44}. All input features and regression targets are standardised using the parameters mean and standard deviation, computed on the training set. The network learning rates are set at $10^{-5}$ and a random seed of 1787452436 is used throughout the experiments. Gaussian noise with a standard deviation 0.1 is added to all input features prior to training. BLSTM-RNN models are trained for a maximum of 100 epochs, however, early stopping is employed where training is stopped when no performance increase (lower SSE) is observed on the validation set after 10 epochs.

Following the training phase, network models are evaluated and selected using validation set CCC \cite{b45}, where higher CCC is better, which is acheived using a forward-pass of the validation set data into the trained network. The CCC measure penalises correlated time-series by applying a penalty of mean-squared error as in \eqref{CCC-def}, where \textit{x} represents predicted values, \textit{y} represents ground-truth values, $\sigma_{xy}$ is the covariance, $\sigma^2$ is the variance and $\mu$ is the mean. Models that perform best during validation set experiments are selected for a final test set pass.

\begin{equation}
  \label{CCC-def}
  CCC = \frac{2 \sigma_{xy}}{\sigma^2_{x} + \sigma^2_{y} + (\mu_{x} - \mu_{y})^2} \
\end{equation}

\section{Initial Results and Discussion}
\subsection{Speech, Head Pose and Eye Gaze Affect Prediction}
This experiment included speech, head pose and eye gaze modalities. New head pose and eye gaze feature sets were generated and these were compared against and combined with speech for continuous affect prediction on the RECOLA \cite{b17} corpus. The head pose and eye gaze feature sets' extraction scripts for this experiment are made available in an accompanying GitHub repository \footnote{\label{myfootnote4}https://github.com/sri-ait-ie/Non-intrusive\_affective\_computing}.
During initial LLD feature exploration, the strongest correlations with both arousal and valence resulted from averages of \textit{x} head location (arousal \textit{r} = -0.26, valence \textit{r} = -0.149) for head pose and \textit{x} gaze angle (arousal \textit{r} = 0.273, valence \textit{r} = 0.245) for eye gaze, both using $W_{s}$ = 8.

The results for the top performing modalities and systems from this experiment are given in Table~\ref{table1}. The top performing arousal system included speech \& head pose while the top performing valence system included speech \& head pose \& eye gaze. There was always a performance increase observed when head pose and/or eye gaze was added to speech for continuous valence prediction and unimodal head pose was shown to perform well for arousal prediction from the non-verbal channel (validation set CCC = 0.535). This experiment demonstrated the early promise of the proposed features as performance improvements above that of unimodal speech were shown to be possible with the proposed sets.

\begin{table}[htbp]
\caption{BLSTM-RNN Results For Top Performing Systems}
\begin{center}
\begin{tabular}{|c|c|c|c|c|}
\hline
\multirow{2}{*}{\textbf{\begin{tabular}{c}System\\(Evaluation)\end{tabular}}}&\multicolumn{2}{|c|}{\textbf{Arousal}}&\multicolumn{2}{|c|}{\textbf{Valence}}\\\cline{2-5} & \textbf{SSE} & \textbf{CCC} & \textbf{SSE} & \textbf{CCC} \\
\hline
\begin{tabular}{c}Speech \& Head Pose\\(Validation)\end{tabular}& 0.152 & \textbf{0.771} & 0.352 & 0.418  \\

\begin{tabular}{c}Speech \& Head Pose\\ \& Eye Gaze (Validation)\end{tabular}& 0.177 & 0.744 & 0.355 & \textbf{0.444} \\

\begin{tabular}{c}Top Performing Validation\\Systems (Test)\end{tabular}& - & 0.779 & - & 0.326 \\
\hline
\end{tabular}
\label{table1}
\end{center}
\end{table}

\subsection{Eye-based Continuous Affect Prediction}
Based on the initial experiment, it was considered that the performance of the eye gaze feature set could be improved with further research. Therefore, a logical next step in this work dealt with further exploration of eye-based cues in order to more fully explore this modality for continuous affect prediction. This experiment was intended to provide further evidence for the use or omission of eye-based input for continuous affect prediction. Pupillometry and direct gaze knowledge measures in addition to eye gaze features previously investigated were evaluated combined and compared with speech on the RECOLA \cite{b17} corpus. An interesting result for the exploration of the final eye-based feature sets for both arousal and valence from these experiments includes the consistent removal of pupil dilation and constriction during feature selection. Also of note is the consistent retention of direct gaze-based features, namely, direct gaze ratio and time in seconds [mean, max, total], across both arousal and valence feature sets after selection.

The final results for this experiment can be seen in Table~\ref{table2}. The results show that eye-based cues, considered with speech, provide performance benefits for continuous affect prediction. This result is an improvement on the previous experiment incorporating eye-based cues in the form of gaze as these results show that there can be a benefit of include eye-based cues with speech for arousal prediction. Unfortunately, adding the eye-based cues to speech for valence prediction did not outperform unimodal speech which indicates that these cues on their own, estimated form video, are ineffective when considered with speech for continuous valence prediction.

\begin{table}[htbp]
\caption{Final BLSTM-RNN Results For Systems Including Speech}
\begin{center}
\begin{tabular}{|c|c|c|c|c|}
\hline
\multirow{2}{*}{\textbf{\begin{tabular}{c}System\\(Evaluation)\end{tabular}}}&\multicolumn{2}{|c|}{\textbf{Arousal}}&\multicolumn{2}{|c|}{\textbf{Valence}}\\\cline{2-5} & \textbf{SSE} & \textbf{CCC} & \textbf{SSE} & \textbf{CCC} \\
\hline
\begin{tabular}{c}Speech-based\\(Validation)\end{tabular}& 0.192 & 0.675 & 0.391 & \textbf{0.103} \\
\begin{tabular}{c}Speech \& Eye-based\\(Validation)\end{tabular}& 0.17 & \textbf{0.737} & 0.402 &  0.059 \\
\begin{tabular}{c}Speech \& Eye-based\\(Test)\end{tabular}& - & 0.72 & - & - \\
\hline
\end{tabular}
\label{table2}
\end{center}
\end{table}

\section{Conclusions and Future Work}
The contributions of this work thus far include hand-crafted feature sets based on head and eye-based cues. Software to extract these sets are provided publicy, for use by the affective computing community. Good results have been achieved for unimodal head pose, and head pose and/or eye-based cues when combined with speech, which shows the benefits of considering these features. A tentative future work plan involves generating speech, head and eye-based cues using automated\footnote{\label{myfootnote5}http://docs.h2o.ai/driverless-ai/latest-stable/docs/userguide/index.html} and CNN-based feature generation on the raw data in order to ensure full exploration of these modalities. These automatically generated/learned features sets will be compared against and combined with the hand-crafted feature sets. Feature selection and fusion will be further investigated, in order to find the optimal feature sets and fusion strategy for said cues. Evaluation of the features on the SEMAINE \cite{b13} data set is also planned to assess feature generalisability. This work will provide the affective computing community with new knowledge and tools for feature extraction from video sequences that can provide for transparency of future results, ease of use for researchers and provide ground-work towards shared standard feature sets in a fashion similar to that of speech \cite{b22}.

\vspace{12pt}
\color{red}

\end{document}